\documentclass{osa-article}

\journal{oe}
\usepackage{siunitx}
\usepackage{amsmath}
\usepackage{amsfonts}
\usepackage{graphicx}
\usepackage[caption=false]{subfig}
\usepackage{dcolumn}
\usepackage{multirow}
\usepackage{hyperref}
\usepackage{xcolor}

\hypersetup{
    colorlinks=true,       % false: boxed links; true: colored links
    linkcolor=blue,          % color of internal links (change box color with linkbordercolor)
    citecolor=blue,        % color of links to bibliography
    filecolor=blue,      % color of file links
    urlcolor=blue           % color of external links
}

\graphicspath{{./}{./Figures/}{./Theory/}}

\begin{document}

\title{\(\Lambda \)-enhanced gray molasses in a tetrahedral laser beam geometry}

\author{D. S. Barker,\authormark{1,2} E. B. Norrgard,\authormark{1}
N. N. Klimov,\authormark{1} J. A. Fedchak,\authormark{1}
J. Scherschligt,\authormark{1} and S. Eckel\authormark{1,3}}

\address{\authormark{1}Sensor Science Division, National Institute of Standards and Technology, Gaithersburg, MD 20899, USA\\
% \authormark{2}Joint Quantum Institute, University of Maryland and National Institute of Standards and Technology, College Park, MD 20742, USA\\
}

\authormark{2}{daniel.barker@nist.gov} %% email address is required
\authormark{3}{stephen.eckel@nist.gov}



\begin{abstract*}
We report observation of sub-Doppler cooling of lithium using an irregular-tetrahedral laser beam arrangement, which is produced by a nanofabricated diffraction grating.
We are able to capture \(11(2)~\si{\percent}\) of the lithium atoms from a grating magneto-optical trap into \(\Lambda \)-enhanced \(D_1\) gray molasses.
The molasses cools the captured atoms to a radial temperature of \(60(9)~\si{\micro\kelvin}\) and an axial temperature of \(23(3)~\si{\micro\kelvin}\).
In contrast to results from conventional counterpropagating beam configurations, we do not observe cooling when our optical fields are detuned from Raman resonance.
An optical Bloch equation simulation of the cooling dynamics agrees with our data.
Our results show that grating magneto-optical traps can serve as a robust source of cold atoms for tweezer-array and atom-chip experiments, even when the atomic species is not amenable to sub-Doppler cooling in bright optical molasses.
\end{abstract*}

%%%%%%%%%%%%%%%%%%%%%%%%%%  body  %%%%%%%%%%%%%%%%%%%%%%%%%%
\section{Introduction}

There is a large and growing effort to produce deployable quantum sensors based on laser-cooled atoms~\cite{Rushton2014, Keil2016}.
An important step toward the goal of fieldable atomic sensors is miniaturizating the mainstay of laser-cooling experiments: the magneto-optical trap (MOT).
One promising approach to MOT miniaturization is the grating MOT, which replaces most of the magneto-optical trap's expansive optical layout with a compact set of diffraction gratings~\cite{Vangeleyn2010,Barker2019}.
Grating MOTs are being integrated into a variety of quantum instruments and sensors, including atomic clocks~\cite{Elvin2019}, atom interferometers~\cite{Lee2021}, electron-beam sources~\cite{Franssen2019}, magnetometers~\cite{McGilligan2017}, and vacuum gauges~\cite{McGilligan2017, Eckel2018}.
Both the diffraction gratings~\cite{Nshii2013, Cotter2016} and beam launch optics~\cite{McGehee2021} of a grating MOT are amenable to nanofabrication.
Grating MOTs have also been integrated with compact vacuum systems~\cite{McGilligan2020, Burrow2021}.
Chip-scale quantum sensors appear attainable in the near future on the grating MOT platform.

The simplicity of the grating MOT optics comes at the expense of decreased symmetry~\cite{Raab1987}.
In a conventional six-beam MOT, formed using three orthogonal pairs of counterpropagating laser beams, the trapping forces are anti-symmetric under inversion.
Thus, many properties, such as trapping forces and equilibrium temperature, are relatively easy to compute at first order by considering three nearly-identical, anti-symmetric, one-dimensional traps.
By contrast, a grating MOT is formed using four or more non-orthogonal laser beams~\cite{Vangeleyn2010}, and correct predictions of trapping properties are only possible by considering the full geometry of the trap~\cite{McGilligan2015, Imhof2017}.
Sub-Doppler cooling in optical molasses becomes particularly complicated because the polarization and intensity gradients arising from the grating beam geometry do not map onto either of the standard lin\(\perp \)lin or \(\sigma^{+}\sigma^{-}\) polarization gradient cooling mechanisms~\cite{Ungar1989, Lett1989, Petsas1994, Lee2013a}.

Nevertheless, sub-Doppler temperatures have been produced with tetrahedral grating beam configurations in bright optical molasses~\cite{Nshii2013, Lee2013a, McGilligan2017} ('bright' refers to an \(F\rightarrow F'=F+1\) transition~\cite{Boiron1995}, where \(F\) (\(F'\)) is the total angular momentum quantum number of the ground (excited) state).
Bright optical molasses is a powerful cooling method and it is a key component of atomic quantum sensors such as clocks~\cite{Elvin2019} and interferometers~\cite{Hauth2013, Wu2017}.
However, bright molasses is not applicable to all laser-coolable systems as it requires resolved hyperfine structure.
Bright molasses is also not ideal for certain applications, such as loading optical tweezer arrays~\cite{Brown2019}.

As grating MOTs are integrated into more quantum sensors and employed to cool more species~\cite{Barker2019, Sitaram2020}, sub-Doppler cooling methods beyond bright optical molasses must be brought to bear.
Two such methods are gray molasses~\cite{Valentin1992, Grynberg1994, Boiron1995} and \(\Lambda \)-enhanced cooling~\cite{Shahriar1993, Marte1994, Weidemuller1994}.
Both methods exploit dark states that appear in \(F\leq F'\) transitions and three-level \(\Lambda \)-systems, respectively, to combine velocity-selective coherent population trapping (VSCPT) with polarization gradient cooling~\cite{Aspect1988, Shahriar1993, Grynberg1994, Weidemuller1994}.
When the cooling light is blue-detuned from the atomic transition, bright states within the ground state manifold are optically pumped into lower energy dark states.
High-velocity atoms in a dark state can non-adiabatically transfer back to a bright state for further cooling, but low-velocity atoms become trapped in the dark state, yielding sub-Doppler temperatures.
For systems with multiple resolved ground-state hyperfine manifolds, \(\Lambda \)-enhanced cooling and gray molasses can be combined to reach even lower temperatures~\cite{Fernandes2012, Grier2013}.
\(\Lambda \)-enhanced gray molasses has three main advantages over bright optical molasses.
First, it is more widely applicable since many atomic and molecular species have well-resolved \(F\leq F'\) transitions~\cite{Fernandes2012, Truppe2017}.
For laser cooled and trapped molecules, \(\Lambda \)-enhanced cooling has emerged as the standard technique for cooling from \(T\simeq 1~\si{\milli\kelvin}\) to \(T\simeq 10~\si{\micro\kelvin}\)~\cite{Truppe2017,Anderegg2018,McCarron2018}.
Second, it shelves atoms in dark states, greatly reducing secondary photon scattering within the atom cloud.
The resulting high atom densities are essential for devices~\cite{Rosi2018}, such as quantum memories and single-photon sources~\cite{Ornelas-Huerta2020}, that rely on strong light-matter coupling.
Third, it uses blue-detuned light, which induces light-assisted collisions that allow high efficiency loading of optical tweezer arrays~\cite{Grunzweig2010, Brown2019}.
Although gray molasses has been studied in four-beam configurations~\cite{Boiron1995}, to our knowledge, neither it nor \(\Lambda \)-enhanced cooling have been implemented in tetrahedral or pyramidal laser beam geometries.

We report our observation of \(\Lambda \)-enhanced sub-Doppler cooling in a irregular-tetrahedral gray molasses produced using a nanofabricated diffraction grating.
Lithium atoms are precooled in a grating MOT and subsequentally transferred into a gray optical molasses operating on the lithium \(D_1\) line.
The gray molasses captures approximately \(10~\si{\percent}\) of the lithium atoms from the MOT and cools them to average temperatures as low as \(50~\si{\micro\kelvin}\), well below the Doppler temperature \(T_D\approx 140~\si{\micro\kelvin}\) for lithium.
Curiously, we do not see incoherent gray molasses cooling when we detune the molasses from Raman resonance~\cite{Grier2013}.
To understand the low capture efficiency and lack of incoherent gray molasses cooling, we simulate the three-dimensional molasses cooling process with the optical Bloch equations.
The simulations suggest that both the capture efficiency and absence of incoherent cooling are features of our experimental procedure, and not of the grating laser beam geometry.
We expect that the capture efficiency could be increased to greater than \(50~\si{\percent}\) using either deeper precooling or higher molasses intensity.
Our results show that sub-Doppler cooling methods beyond bright optical molasses can operate in the nonorthogonal beam geometries of grating MOTs.

\section{Apparatus}

Our experiments take place within the grating MOT apparatus described in Ref.~\cite{Barker2019}.
Additional technical information about certain aspects of our setup can be found in Refs.~\cite{Starkey2013, Norrgard2018, Fedchak2018, Barker2019a}.
The apparatus has four principal coaxial components: a set of electromagnets, a nanofabricated diffraction grating chip, an input laser beam, and an effusive Li dispenser.
The common axis of the components defines the axial unit vector \(\hat{z}\).
The electromagnets create a quadrupole magnetic field in front of the grating chip and continuously deform that field toward the square-root profile of a Zeeman slower behind the chip~\cite{Phillips1982}.
The diffraction grating chip has three linear diffraction gratings, which are arranged so that their grooves form equilateral triangles.
We define the radial unit vectors \(\hat{x}\) and \(\hat{y}\) to be perpendicular and parallel to the grooves of one of the linear gratings.
At \(\lambda_{\text{Li}}\approx 671~\si{\nano\meter}\), each grating has a first-order diffraction efficiency of \(37(1)~\si{\percent}\) and its Stokes parameters for normally incident, left-hand circularly polarized light are \(Q = 0.16(1)\), \(U = -0.37(1)\), \(V = 0.92(1)\), where \(Q = 1\) (\(Q = -1\)) corresponds to \(p\) (\(s\)) polarization defined relative to the plane of incidence for each linear grating~\cite{gratingnote}.
(Here, and throughout the paper paranthetical quantities represent the standard error).
The input laser beam is
%collinear with the \(z\) axis and 
normally incident to the grating chip, generating six diffracted laser beams at an angle \(\theta_{d} \approx \pm 42~\si{\degree}\) relative to the chip normal \(-\hat{z}\).
The diffracted beams that propagate toward the center of the quadrupole magnetic field combine with the input beam to produce an irregular-tetrahedral laser beam geometry suitable for magneto-optical trapping or optical molasses~\cite{Shimizu1991, Lin1991, Nshii2013, Lee2013a}.
The center of the input laser beam passes through a triangular aperture etched in the grating chip and strikes the Li dispenser.
Lithium atoms emitted from the dispenser are Zeeman slowed behind the chip and then captured into the grating MOT after transiting the chip aperture.

Magneto-optical trapping and \(\Lambda \)-enhanced molasses cooling require distinct input laser beams: the cooling beam and molasses beam, respectively.
Both beams have a \(1/e^2\) radius of approximately \(18~\si{\milli\meter}\) and are stopped to fit the \(22~\si{\milli\meter}\) diameter of the grating by an iris.
The center frequency of the cooling beam is detuned by \(\Delta_{c}\) from the \(^2\text{S}_{1/2}\,(F=2)\rightarrow \, ^2\text{P}_{3/2}\,(F'=3)\) cycling transition.
An electro-optic modulator (EOM) frequency modulates the cooling beam at approximately \(813~\si{\mega\hertz}\), so the \(+1\)st-order sideband is detuned by \(\Delta_{c}\) from the \(^2\text{S}_{1/2}\,(F=1)\rightarrow \, ^2\text{P}_{3/2}\,(F'=2)\) ``repump'' transition.
We set the modulation depth of the EOM to produce a \(1:2\) ratio of the repump intensity to the cooling intensity \(I_c\).
The center frequency of the molasses beam is detuned by \(\Delta_{2}\) from the \(^2\text{S}_{1/2}\,(F=2)\rightarrow \, ^2\text{P}_{1/2}\,(F'=2)\) transition.
Another EOM, nominally operating at the \(^7\)Li ground state hyperfine splitting \(\nu_{hfs}\approx 803.5~\si{\mega\hertz}\), adds sidebands for repumping to the molasses beam and detunes the \(+1\)st-order sideband by \(\Delta_{1}\) from the \(^2\text{S}_{1/2}\,(F=1)\rightarrow \, ^2\text{P}_{1/2}\,(F'=2)\) transition.
The modulation frequency of the molasses EOM controls the \(2\)-photon Raman detuning \(\delta = \Delta_{1}-\Delta_{2}\) of the \(\Lambda \)-system defined by \(^2\text{S}_{1/2}\,(F=1)\), \(^2\text{S}_{1/2}\,(F=2)\), and \(^2\text{P}_{1/2}\,(F'=2)\).
The modulation depth of the molasses EOM governs the relative intensity of the molasses carrier \(I_2\) and the molasses sideband \(I_1\).
The \(^2\text{S}_{1/2}\,\rightarrow \,^2\text{P}_{3/2}\) cooling transition and the \(^2\text{S}_{1/2}\,\rightarrow \,^2\text{P}_{1/2}\) molasses transition both have natural linewidth \(\Gamma_\text{Li}\approx2\pi\times5.87~\si{\mega\hertz}\).
The saturation intensity of the cooling transition is \(I_\text{sat}\approx2.54~\si{\milli\watt\per\centi\meter\squared}\).
The molasses and cooling beams are combined using a polarizing beam cube and a Pockels cell, so both beams are left-hand circularly polarized (\(\sigma^{-}\)) with respect to the quantization axis \(\hat{z}\).

We have implemented three upgrades to the apparatus in Ref.~\cite{Barker2019}.
First, three pairs of magnetic shim coils now surround the vacuum chamber.
The shim coils null the ambient magnetic field and are active throughout our experimental sequence.
Second, the commercial Li vapor source has been replaced with a lower-outgassing, three-dimensionally printed titanium dispenser~\cite{Norrgard2018}.
Third, the dispenser is now placed approximately \(2~\si{\centi\meter}\) behind the nanofabricated grating chip, near the peak magnetic field of the grating MOT's integrated Zeeman slower.
Combined with the reduced \(1/e^2\) radius of the input cooling beam~\cite{Barker2019}, the latter two upgrades have increased the maximum number of trapped lithium atoms by a factor of \(7\).
The higher trapped atom number increased the signal-to-noise ratio of our resonant absorption imaging, allowing us to observe sub-Doppler cooling.

\section{Experiment}

We begin our measurements by preparing a cloud of cold \(^7\)Li atoms in the grating MOT.\@
To load atoms, we operate the MOT at a nominal axial magnetic field gradient \(B'=6~\si{\milli\tesla\per\centi\meter}\), cooling laser detuning \(\Delta_c/\Gamma_\text{Li}=-5.1\), and carrier saturation parameter \(s_c=I_c/I_\text{sat}=4.9\).
The MOT loading stage lasts for \(0.65~\si{\second}\), which is similar to the trap lifetime \(\tau\approx0.8~\si{\second}\).
We then compress the MOT for \(2.5~\si{\milli\second}\) by jumping the cooling laser detuning to \(\Delta_c/\Gamma_\text{Li}=-2.0\) and simultaneously reducing the carrier saturation parameter to \(s_c=0.5\).
After compression, the MOT contains approximately \(7\times10^6\)~\(^7\)Li atoms at a radial temperature \(T_x\approx650~\si{\micro\kelvin}\) and an axial temperature \(T_z\approx350~\si{\micro\kelvin}\).

We transfer the atoms from the MOT into the \(\Lambda \)-enhanced gray molasses by extinguishing the cooling beam, turning off the MOT electromagnets, and activating the molasses beam.
The nominal parameters of the molasses beam are \(1\)-photon detuning \(\Delta_2/\Gamma_\text{Li}=3.1\), \(2\)-photon detuning \(\delta/\Gamma_\text{Li}=0\), total saturation \(s_m=(I_1+I_2)/I_\text{sat}=3.2(1)\), and sideband-to-carrier ratio \(I_1/I_2=0.233(4)\).
The momentum distribution of the atom cloud evolves within the gray molasses for \(1~\si{\milli\second}\).

\begin{figure}[t]
\centering\includegraphics[width=\textwidth]{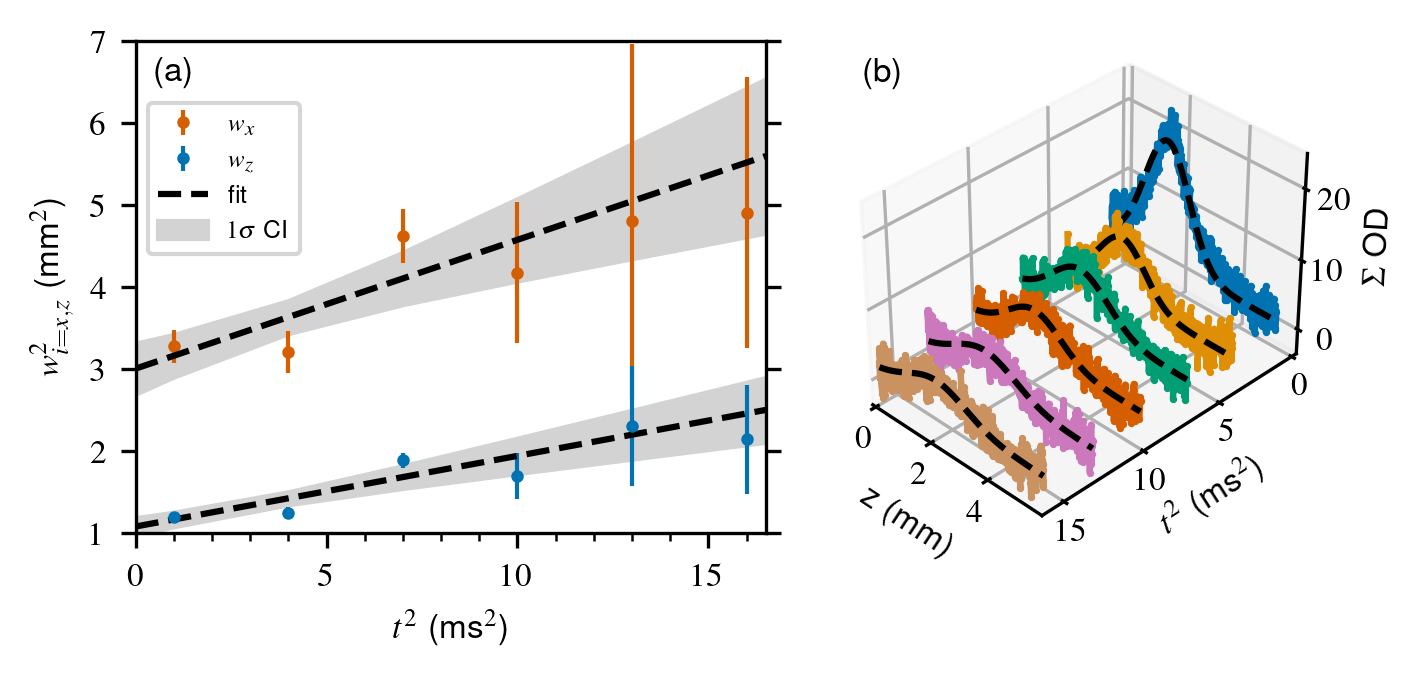}
\caption{\label{fig:tof}
  Example time-of-flight sequence.
  (a) The \(x\) (red) and \(z\) (blue) squared \(1/e\) radii \(w^{2}_{i=x,z}(t)\) plotted as a function of \(t^2\).
  Errorbars, some of which are smaller than the data points, denote the standard error.
  Black dashed lines are fits to Eq.~\ref{eq:tof} with the associated \(1\sigma \) confidence interval (CI) shown in gray.
  For this sequence, \(T_x = 66(26)~\si{\micro\kelvin}\) and \(T_z = 36(11)~\si{\micro\kelvin}\).
  (b) Average absorption images integrated over \(x\) (colored points) at each time of flight for the sequence in (a).
  Black dashed lines show integrated \(2\)-dimensional Gaussian fits to the average absorption image.
  }
\end{figure}

We measure the temperature of the atoms with a time of flight method.
After shutting off the molasses beam, we record the distribution of the atom cloud using resonant absorption imaging for a sequence of times of flight \(t\).
The absorption images are fit to a \(2\)-dimensional Gaussian to extract the radial and axial \(1/e\) widths of the atom cloud \(w_x(t)\) and \(w_z(t)\), respectively.
We repeat the time-of-flight sequence four times and fit the average atom cloud radii to
\begin{equation}
  \label{eq:tof}
  w^{2}_{i}(t)=w^{2}_{i}(0)+\frac{2 k_B T_i}{m}t^{2},
\end{equation}
where \(k_B\) is Boltzmann's constant, \(m\) is the atomic mass of \(^7\)Li, and \(T_i\) is the molasses temperature along axis \(i=x,z\).
Figure~\ref{fig:tof} shows an example time-of-flight sequence for the nominal molasses operation parameters.
Before scanning a molasses parameter, we run a time-of-flight sequence with the nominal parameters to ensure that the shim coils are properly nulling residual magnetic fields.

\begin{figure}[t]
\centering\includegraphics[width=\textwidth]{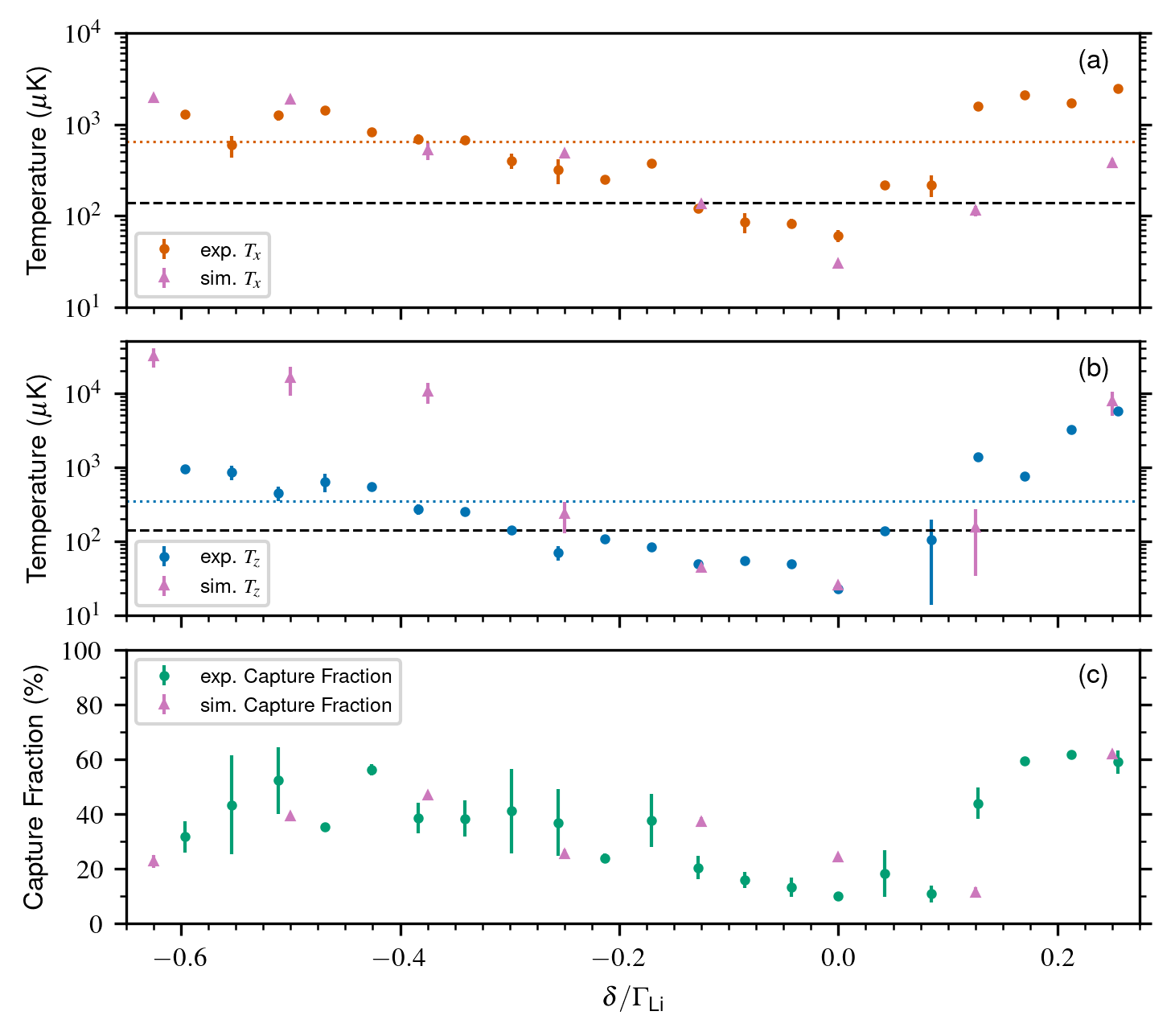}
\caption{\label{fig:delta_molasses}
  \(\Lambda \)-enhanced molasses performance as a function of Raman detuning \(\delta/\Gamma_\text{Li} \).
  (a) and (b) show the measured radial temperature \(T_x\) (red circles) and axial temperature \(T_z\) (blue circles), respectively.
  The horizontal dotted lines indicate the temperature of the compressed grating MOT and the horizontal dashed lines represent the Doppler cooling limit for the lithium.
  (c) shows the capture efficiency of the \(\Lambda \)-enhanced molasses (green circles).
  In all three subplots, purple triangles are results of optical Bloch equation simulations of the \(\Lambda \)-enhanced cooling process (see Sec.~\ref{sec:theory}).
  Errorbars represent the standard error and are often smaller than the data points.
  }
\end{figure}

We study the performance of the \(\Lambda \)-enhanced gray molasses as a function of \(\delta \), \(\Delta_2\), and \(I_1/I_2\).
Figure~\ref{fig:delta_molasses} shows the atom cloud temperatures \(T_x\) and \(T_z\), as well as the the molasses capture efficiency, as a function of \(\delta/\Gamma_{\text{Li}} \).
For the data in Fig.~\ref{fig:delta_molasses}, \(\Delta_2/\Gamma_\text{Li}=3.1\), \(s_m=3.2(1)\), and \(I_1/I_2=0.233(4)\).
We observe strong sub-Doppler cooling to \(T_x=60(9)~\si{\micro\kelvin}\) and \(T_z=23(3)~\si{\micro\kelvin}\) at Raman resonance (\(\delta/\Gamma_\text{Li}=0\)), which indicates that \(\Lambda \)-enhanced cooling occurs in our tetrahedral laser beam geometry.
Our final temperatures are comparable to those achieved in conventional \(6\)-beam \(\Lambda \)-enhanced molasses configurations with lithium~\cite{Grier2013, Burchianti2014, Sievers2015, Satter2018, Kim2019}.
For \(\delta/\Gamma_\text{Li}>0\), the \(\Lambda \)-enhanced molasses heats the atom cloud as expected when \(I_2 > I_1\)~\cite{Sievers2015}.
However, in contrast to prior experiments, we see no evidence of incoherent gray molasses cooling when \(\delta/\Gamma_\text{Li}<0\)~\cite{Fernandes2012, Grier2013, Salomon2013, Nath2013, Burchianti2014, Sievers2015, Bouton2015, Chen2016, Colzi2016, Satter2018, Kim2019}.
The fraction of atoms that are captured from our MOT into the molasses stage at Raman resonance is \(11(2)~\si{\percent}\), substantially lower than the capture efficiency for most \(6\)-beam lithium experiments using molasses with similar intensity~\cite{Burchianti2014, Sievers2015, Satter2018}.
Our data suggest that incoherent \(D_1\) gray molasses cooling is absent for our molasses parameters in the non-orthogonal beam arrangements produced by diffraction grating chips (see Sec.~\ref{sec:theory}).

\begin{figure}[t]
\centering\includegraphics[width=\textwidth]{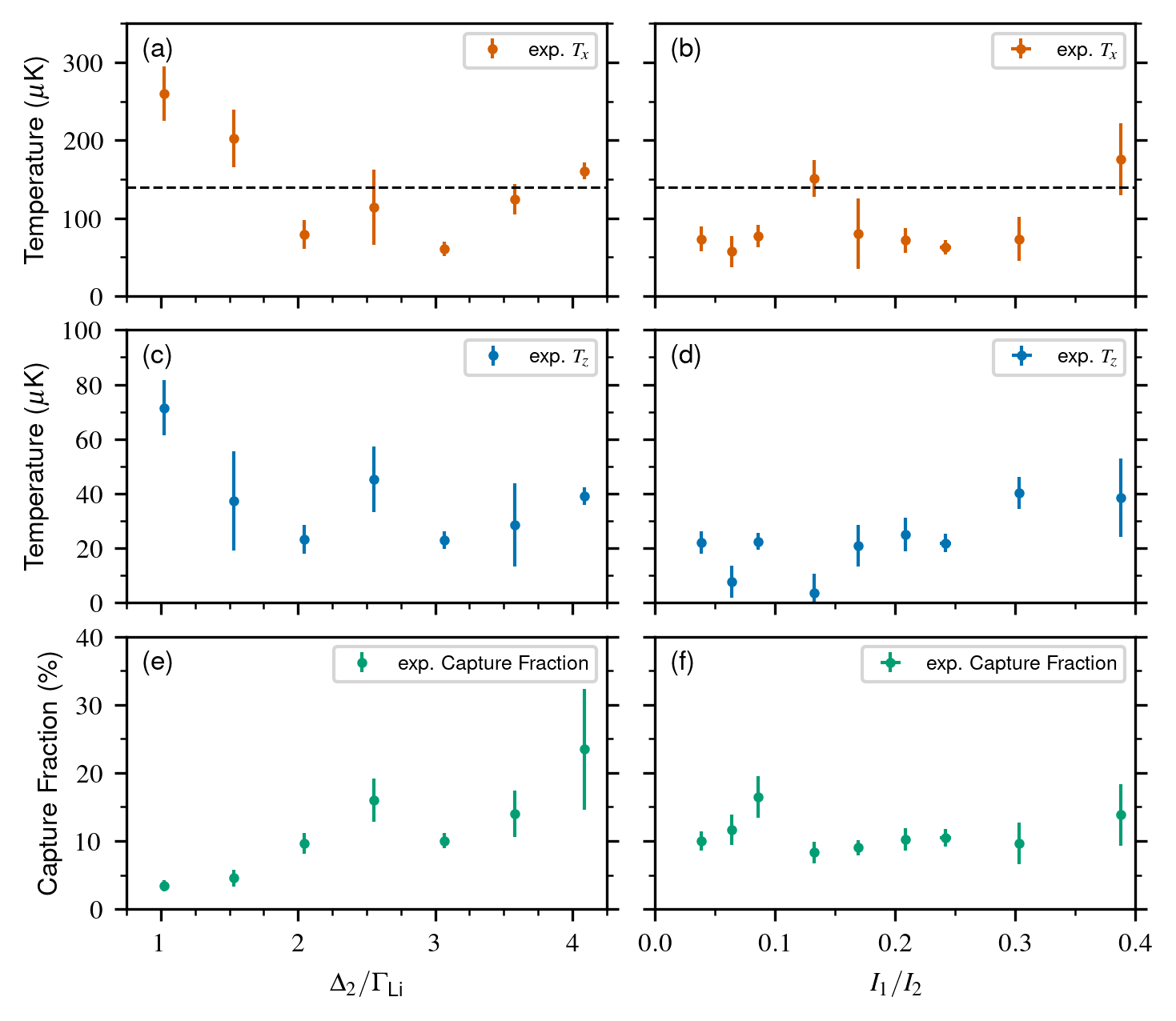}
\caption{\label{fig:dii_molasses}
  \(\Lambda \)-enhanced molasses performance as a function of \(1\)-photon detuning \(\Delta_2 \) and sideband-to-carrier ratio \(I_1/I_2\).
  (a), (c), (e) show \(T_x\) (red circles), \(T_z\) (blue circles), and the capture efficiency (green circles) as a function of \(\Delta_2 \), respectively.
  (b), (d), (f) are the corresponding plots of the molasses performance as a function of \(I_1/I_2\).
  The horizontal dashed lines in (a) and (b) indicate the Doppler cooling limit for lithium.
  In all subplots, errorbars show the standard error.
  }
\end{figure}

Figure~\ref{fig:dii_molasses} shows the change \(\Lambda \)-enhanced molasses temperature and capture efficiency with \(\Delta_2\) and \(I_1/I_2\).
The data in the left column of Fig.~\ref{fig:dii_molasses} has \(s_m=3.2(1)\) and \(I_1/I_2=0.233(4)\), while the data in the right column of Fig.~\ref{fig:dii_molasses} has \(\Delta_2/\Gamma_\text{Li}=3.1\).
We note that, because optical power in the \(-1\)st-order molasses beam sideband is wasted, \(s_m\) decreases with \(I_1/I_2\) in the right column of Fig.~\ref{fig:dii_molasses}.
For both data sets in Fig.~\ref{fig:dii_molasses}, the Raman detuning \(\delta/\Gamma_\text{Li}=0\).
The radial and axial temperatures exhibit a shallow minimum between \(\Delta_2/\Gamma_\text{Li} = 2.0\) and \(\Delta_2/\Gamma_\text{Li} = 3.0\), which is smaller than the detuning \(\Delta_2/\Gamma \approx 5\) of the temperature minimum in most \(6\)-beam experiments~\cite{Fernandes2012, Salomon2013, Nath2013, Burchianti2014, Chen2016, Colzi2016, Satter2018}.
Our molasses capture efficiency increases with \(\Delta_2\), but more slowly than in the \(6\)-beam geometry~\cite{Fernandes2012, Burchianti2014, Sievers2015, Chen2016, Colzi2016, Satter2018}.
The molasses performance as a function of \(I_1/I_2\) is also consistent with \(6\)-beam experiments~\cite{Grier2013, Nath2013, Colzi2016, Satter2018}: the temperature increases with \(I_1/I_2\) and the capture efficiency is flat.

\section{\label{sec:theory} Simulation}

To investigate the lack of incoherent gray molasses and low capture efficiency of the tetrahedral gray molasses (see Fig.~\ref{fig:delta_molasses}), we simulate the molasses cooling process using the optical Bloch equations.
Our simulation numerically integrates the classical motion and density matrix evolution of the atom in the optical field of the tetrahedral \(\Lambda \)-enhanced gray molasses for \(1~\si{\milli\second}\).
We generate the optical Bloch equations for the Zeeman level structure of \(^2\text{S}_{1/2}(F=1)\), \(^2\text{S}_{1/2}(F=2)\), and \(^2\text{P}_{1/2}(F'=2)\) with our grating laser beam arrangement using the {\tt pylcp} Python package~\cite{Eckel2020}.
We program the laser beams with pure circular polarization and uniform intensity profiles~\cite{polnote}, but we reduce the intensity of the diffracted beams to account for the Gaussian intensity profile of the input molasses beam projected onto the measured position of the MOT.
Each simulated trajectory initializes with an atom at the origin, with a random velocity drawn from a Maxwell-Boltzmann distribution, and a density matrix with equal populations in each of the \(^2\text{S}_{1/2}\,(F=1)\) Zeeman states.  
The initial density matrix ensures atoms are not initialized in a coherent dark state.
%\textcolor{red}{Each simulation trajectory begins by sampling an atom at the origin from a Maxwell-Boltzmann velocity distribution.
%We initialize the atomic density matrix with an equal population in each of the \(^2\text{S}_{1/2}\,(F=1)\) Zeeman states and all other density matrix elements equal to zero.
%We chose this initial density matrix because it ensures that the atom is not initialized in a coherent dark state.}
%Our simulation numerically integrates the classical motion and density matrix evolution of the atom in the optical field of the tetrahedral \(\Lambda \)-enhanced gray molasses for \(1~\si{\milli\second}\).
The simulation also includes the effects of gravity and random recoil due to spontaneous emission.

%Need a new paragraph, was too long
%The trajectory simulations include gravitational acceleration and random recoil effects due to spontaneous photon emission, but neglects secondary scattering processes, quantum jumps, and stimulated momentum diffusion.
Both \(\Lambda \)-enhanced cooling and gray molasses are sub-recoil cooling techniques~\cite{Shahriar1993, Grynberg1994, Marte1994, Weidemuller1994}.
%In the absence of 
Because our simulation does not include the effects of secondary scattering, stimulated momentum diffusion, quantum jumps and technical noise, the simulations produce temperatures far lower than our experimental observations.
Secondary scattering is insignificant in our experiment because the radial and axial velocity distributions are not in thermal equilibrium (see Fig.~\ref{fig:delta_molasses}).
The stimulated momentum diffusion rate is not calculated by the {\tt pylcp} package because there is not a closed-form expression for it as function of velocity and approximations based on two-level atoms are not applicable to our degenerate \(\Lambda \) level structure~\cite{Ungar1989, Ashkin1980, Eckel2020}.
Quantum jumps would heat the simulated molasses since they disturb the evolution of the density matrix into a coherent dark state.
However, prior \(\Lambda \)-enhanced cooling simulations that included quantum jumps also did not quantitatively reproduce experimentally observed temperatures~\cite{Sievers2015, Bruce2017}.
We apply a uniform magnetic field \(B_z=40~\si{\micro\tesla}\) along the axial direction \(\hat{z}\) in our simulations to approximate the effect of technical noise sources (such as residual magnetic field gradients and Raman detuning jitter due to phase noise from the frequency synthesizer driving the molasses EOM)~\cite{techsimnote}.
We chose the value of \(B_z\) so the simulated and experimental axial temperature at \(\delta/\Gamma_\text{Li}=0\) would match after \(200\) trajectories.
%; it also approximates heating due quantum jumps and stimulated emission. 

\begin{figure}[t]
\centering\includegraphics[width=\textwidth]{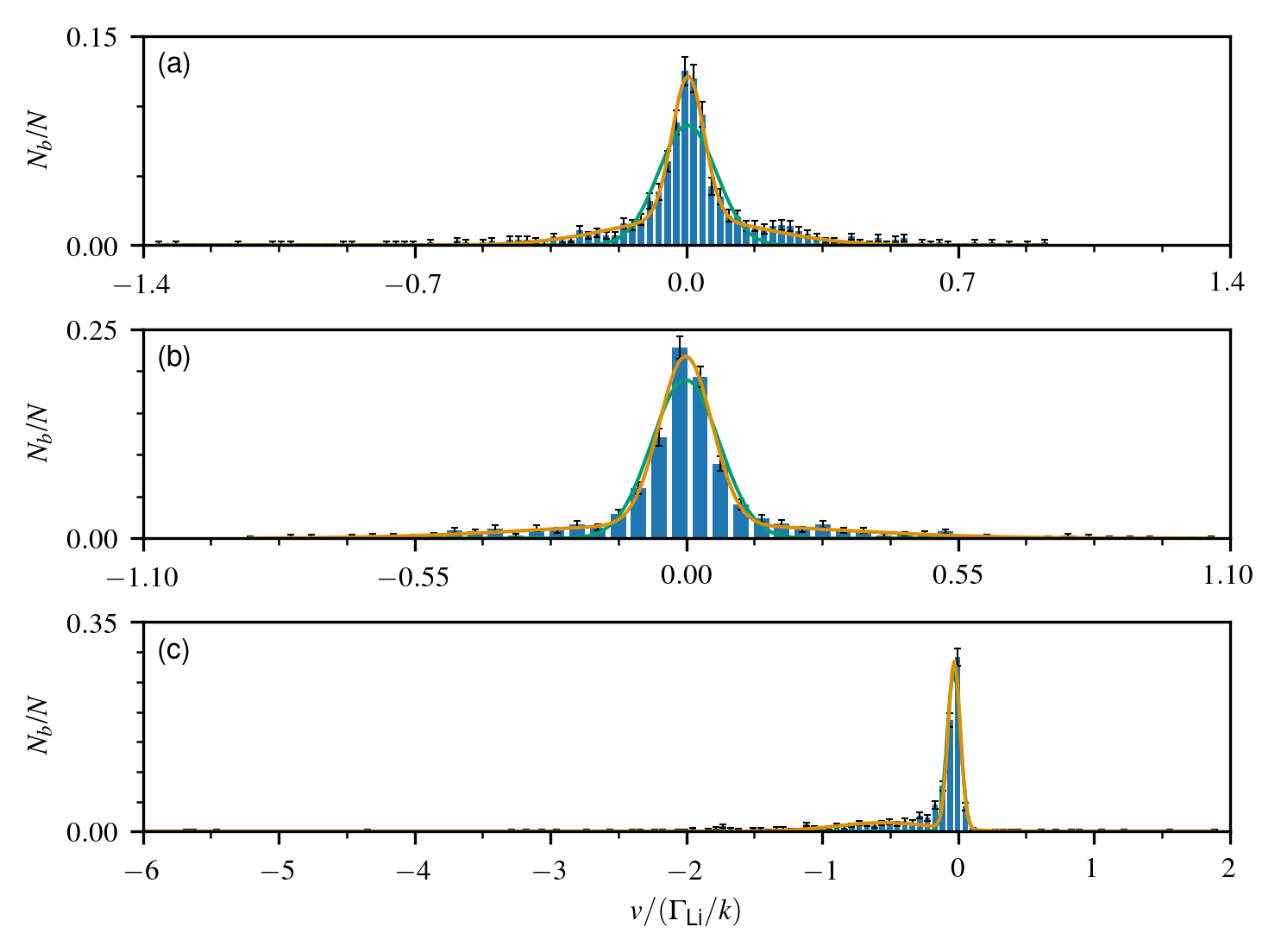}
\caption{\label{fig:simfit}
  Example simulated velocity distribution of \(\Lambda \)-enhanced gray molasses in our tetrahedral laser beam geometry.
  The fraction of atoms in each velocity bin \(N_b/N\) (blue pillars) along the \(x\), \(y\), and \(z\) axes are shown as a function of velocity \(v\) in subplot (a), (b), and (c), respectively.
  Black errorbars on each pillar are the uncertainty in \(N_b/N\), calculated following Ref.~\cite{Knuth2019}.
  We fit the velocity distributions along each axis to a unimodal Gaussian distribution (green line) and a bimodal Gaussian distribution (orange line) to extract the temperature.
  The example velocity distributions were constructed from \(1000\) simulated molasses cooling trajectories with \(\Delta_2/\Gamma_\text{Li}=3.0\), \(\delta/\Gamma_\text{Li}=0\), \(s_m=3.2\), and \(I_1/I_2=0.24\).
  The wavenumber of the molasses cooling transition is \(k=2\pi/\lambda_\text{Li}\).}
\end{figure}

We integrate \(1000\) atomic trajectories within the molasses at each of \(8\) Raman detunings.
The other molasses parameters for the simulated trajectories are \(\Delta_2/\Gamma_\text{Li}=3.0\), \(s_m=3.2\), and \(I_1/I_2=0.24\); chosen to match the experimental conditions for Fig.~\ref{fig:delta_molasses}.
To increase the number of trajectories that the molasses captures, we sample the initial atomic velocities from an isotropic Maxwell-Boltzmann distribution at the Doppler temperature.
We construct a histogram of the final velocities along each axis following a Bayesian approach~\cite{Knuth2019}.
Each histogram is fit with both a unimodal and a bimodal Gaussian distribution to extract the molasses temperature, which is the smaller temperature in the case of the bimodal distribution.
To mitigate overfitting, we use the Bayesian information criterion to decide whether the unimodal or bimodal fit best represents the binned trajectories~\cite{Kass1995, Burnham2004, bicnote}.
Fig.~\ref{fig:simfit} shows the histogrammed simulation results and Gaussian fits for \(\delta/\Gamma_\text{Li} = 0\).
The asymmetry of the tetrahedral laser arrangement is most apparent in the velocity distribution along \(\hat{z}\).
To compare with the experimental data, we average the simulated temperatures along \(\hat{x}\) and \(\hat{y}\) to compute the simulated radial molasses temperature.
We plot the simulated radial and axial molasses temperatures as purple triangles in Fig.~\ref{fig:delta_molasses}(a) and Fig.~\ref{fig:delta_molasses}(b), respectively. %chktex 36
The simulations reproduce the main features of the data: sub-Doppler cooling when \(\delta/\Gamma_\text{Li}\approx 0\) and strong heating when \(\delta/\Gamma_{\text{Li}}\lesssim -0.25\) or \(\delta/\Gamma_{\text{Li}}\gtrsim 0.1\).
The comparatively large disagreement between the axial simulation and experiment near \(\delta/\Gamma_{\text{Li}}\approx -0.5\) may be due to the highest velocity atoms escaping the field of view of our imaging system.

We simulate the capture efficiency of the molasses by integrating \(200\) trajectories with initial velocities sampled from an anisotropic Maxwell-Boltzmann distribution with radial and axial temperatures matching the compressed MOT.\@
Once again, the simulated molasses parameters are similar to the experiment: \(\Delta_2/\Gamma_\text{Li}=3.0\), \(s_m=3.2\), and \(I_1/I_2=0.24\).
After \(1~\si{\milli\second}\) of integration, we construct empirical cumulative distribution functions (ECDF) for the final velocities along each axis.
(Histograms of the final velocities do not reliably extract the sub-Doppler features for this simulation set).
Single, double, and triple Gaussian error functions are fit to each ECDF with the Bayesian information criterion determining the best model for the simulated data.
The triple error function is only fit to the axial velocities, which are not fully described by a bimodal distribution when the capture efficiency is low.
Most simulated trajectories escape the molasses along the axial direction \(\hat{z}\), so we compute the simulated molasses capture efficiency using the amplitude of the coldest error function from the fit to the axial velocities.
Purple triangles in Fig.~\ref{fig:delta_molasses}(c) show the simulated capture efficiency as a function of \(\delta/\Gamma_{\text{Li}} \). % chktex 36
The simulations predict a higher capture efficiency at Raman resonance, but otherwise agree with the experimental data.
The trajectory set that we initialized at the Doppler temperature had a capture efficiency of approximately \(60~\si{\percent}\) at \(\delta/\Gamma_\text{Li}=0\) (see Fig.~\ref{fig:simfit}), so the small experimentally observed capture efficiency is not inherent to the tetrahedral laser beam arrangement.
The number of atoms in the molasses would be improved with better precooling in the grating MOT or, possibly, by increasing the molasses intensity \(s_m\).

\begin{figure}[t]
\centering\includegraphics[width=\textwidth]{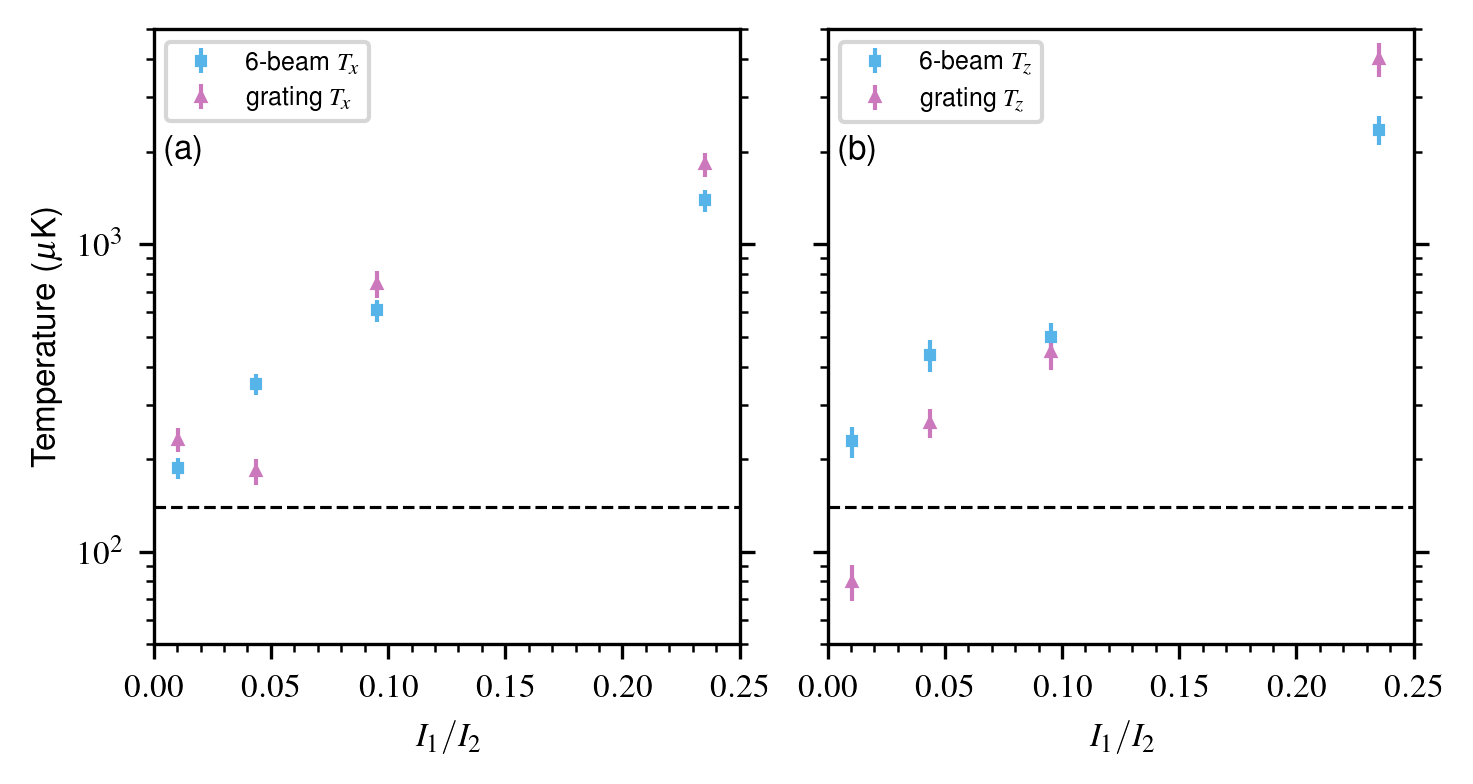}
\caption{\label{fig:sb_sim}
  Simulated temperature of \(\Lambda \)-enhanced gray molasses as a function of \(I_1/I_2\).
  (a) shows simulated radial temperatures and (b) shows simulated axial temperatures.
  Purple triangles and blue squares show simulation results for the grating beam configuration and a conventional \(6\)-beam configuration, respectively.
  The horizontal dashed lines indicate the Doppler cooling limit for lithium.
  The grating molasses and the \(6\)-beam molasses have the same total intensity.
  Each temperature was extracted from \(200\) molasses cooling trajectories with \(\Delta_2/\Gamma_\text{Li}=3.0\), \(\delta/\Gamma_\text{Li}=-0.5\), and \(B_z=0~\si{\micro\tesla}\).
  At \(I_1/I_2=0.24\), \(s_m=3.2\).
  In both subplots, errorbars show the standard error.
  }
\end{figure}

The apparent lack of incoherent gray molasses cooling in the tetrahedral beam geometry is due to our choice of experimental parameters.
When \(\delta/\Gamma_\text{Li} < 0\), atoms in \(^2\text{S}_{1/2}\,(F=1)\) Zeeman states gain kinetic energy whenever they are optically pumped into \(^2\text{S}_{1/2}\,(F=2)\).
Gray molasses cooling still occurs because the \(^2\text{S}_{1/2}\,(F=2)\rightarrow \, ^2\text{P}_{1/2}\,(F'=2)\) transition still has a dark state.
However, the gray molasses cooling rate only exceeds the heating rate due to \(^2\text{S}_{1/2}\,(F=1)\rightarrow \, ^2\text{S}_{1/2}\,(F=2)\) optical pumping when \(I_1\ll I_2\).
Figure~\ref{fig:sb_sim} shows the simulated \(I_1/I_2\) dependence of the gray molasses radial and axial temperatures at \(\delta/\Gamma_{\text{Li}}=-0.5\).
Each temperature was extracted from a Gaussian fit to the binned final velocities of \(200\) trajectories initialized at the Doppler temperature.
To avoid washing out the sub-Doppler feature, \(B_z=0~\si{\micro\tesla}\) for this simulation set.
The simulations of the grating beam geometry indicate that incoherent molasses temperatures near the Doppler limit are achievable when \(I_1/I_2<0.05\), comparable to both measurements in \(6\)-beam experiments~\cite{Grier2013, Kim2019} and our own simulations of \(6\)-beam gray molasses cooling (see Fig.~\ref{fig:sb_sim}).

\section{Conclusion}

We have demonstrated \(\Lambda \)-enhanced gray molasses cooling in a tetrahedral beam configuration.
A single input laser beam striking nanofabricated grating chip produces the molasses light field.
The \(\Lambda \)-enhanced molasses cools \(^{7}\)Li atoms to radial and axial temperatures as low as \(T_x=60(9)~\si{\micro\kelvin}\) and \(T_z=23(3)~\si{\micro\kelvin}\), respectively.
The molasses captures \(11(2)~\si{\percent}\) of the lithium atoms from a grating MOT.\@
Optical Bloch equation simulations of the \(\Lambda \)-enhanced cooling process adequately reproduce our measured temperature and capture efficiency.
The simulations suggest that the capture efficiency of the molasses can be improved by either increasing the molasses intensity or reducing the temperature of the compressed grating MOT.\@
The simulations also indicate that sub-Doppler temperatures can be reached with a small red detuning from Raman resonance, which maximizes the gray molasses capture efficiency~\cite{Burchianti2014, Colzi2016, Satter2018}.
Our work demonstrates that sub-Doppler temperatures are attainable in the beam geometries produced by diffraction grating chips, even for atomic species or molecules that are not amenable to bright molasses cooling.
The realization of dark-state-based laser cooling with a diffraction grating chip opens new application spaces for grating MOTs, allowing them to serve as atom sources for quantum network nodes~\cite{Ornelas-Huerta2020}, primary vacuum gauges~\cite{Eckel2018}, and Rydberg-atom quantum computers~\cite{Brown2019}.

\section*{Funding}

We acknowledge funding from the National Institute of Standards and Technology.

\section*{Acknowledgments}

We thank T. Bui and R. Fasano for their careful reading of the manuscript.
We also thank J. Fletcher for assistance assembling the molasses laser system and T. Koch for bringing Ref.~\cite{Knuth2019} to our attention.

\section*{Disclosures}

The authors declare no conflict of interest.

%%%%%%%%%% If using BibTeX:
\bibliography{lambda_molasses}

\begin{thebibliography}{10}
\newcommand{\enquote}[1]{``#1''}

\bibitem{Rushton2014}
J.~A. Rushton, M.~Aldous, and M.~D. Himsworth, \enquote{{Contributed Review:
  The feasibility of a fully miniaturized magneto-optical trap for portable
  ultracold quantum technology},} {\protect\JournalTitle{Review of Scientific
  Instruments}} \textbf{85}, 121501 (2014).

\bibitem{Keil2016}
M.~Keil, O.~Amit, S.~Zhou, D.~Groswasser, Y.~Japha, and R.~Folman,
  \enquote{{Fifteen years of cold matter on the atom chip: promise,
  realizations, and prospects},} {\protect\JournalTitle{Journal of Modern
  Optics}} \textbf{63}, 1840 (2016).

\bibitem{Vangeleyn2010}
M.~Vangeleyn, P.~F. Griffin, E.~Riis, and A.~S. Arnold, \enquote{{Laser cooling
  with a single laser beam and a planar diffractor},}
  {\protect\JournalTitle{Optics Letters}} \textbf{35}, 3453 (2010).

\bibitem{Barker2019}
D.~S. Barker, E.~B. Norrgard, N.~N. Klimov, J.~A. Fedchak, J.~Scherschligt, and
  S.~Eckel, \enquote{{Single-Beam Zeeman Slower and Magneto-Optical Trap Using
  a Nanofabricated Grating},} {\protect\JournalTitle{Physical Review Applied}}
  \textbf{11}, 064023 (2019).

\bibitem{Elvin2019}
R.~Elvin, G.~W. Hoth, M.~Wright, B.~Lewis, J.~P. McGilligan, A.~S. Arnold,
  P.~F. Griffin, and E.~Riis, \enquote{{A cold-atom clock based on a
  diffractive optic},} {\protect\JournalTitle{Optics Express}} \textbf{27},
  38359 (2019).

\bibitem{Lee2021}
J.~Lee, R.~Ding, J.~Christensen, R.~R. Rosenthal, A.~Ison, D.~P. Gillund,
  D.~Bossert, K.~H. Fuerschbach, W.~Kindel, P.~S. Finnegan, J.~R. Wendt,
  M.~Gehl, H.~McGuinness, C.~A. Walker, A.~Lentine, S.~A. Kemme, G.~Biedermann,
  and P.~D.~D. Schwindt, \enquote{{A Cold-Atom Interferometer with
  Microfabricated Gratings and a Single Seed Laser},} ArXiv:2107.04792 (2021).

\bibitem{Franssen2019}
J.~G.~H. Franssen, T.~C.~H. de~Raadt, M.~A.~W. van Ninhuijs, and O.~J. Luiten,
  \enquote{{Compact ultracold electron source based on a grating magneto
  optical trap},} {\protect\JournalTitle{Physical Review Accelerators and
  Beams}} \textbf{22}, 023401 (2019).

\bibitem{McGilligan2017}
J.~P. McGilligan, P.~F. Griffin, R.~Elvin, S.~J. Ingleby, E.~Riis, and A.~S.
  Arnold, \enquote{{Grating chips for quantum technologies},}
  {\protect\JournalTitle{Scientific Reports}} \textbf{7}, 384 (2017).

\bibitem{Eckel2018}
S.~Eckel, D.~S. Barker, J.~A. Fedchak, N.~N. Klimov, E.~Norrgard,
  J.~Scherschligt, C.~Makrides, and E.~Tiesinga, \enquote{{Challenges to
  miniaturizing cold atom technology for deployable vacuum metrology},}
  {\protect\JournalTitle{Metrologia}} \textbf{55}, S182 (2018).

\bibitem{Nshii2013}
C.~C. Nshii, M.~Vangeleyn, J.~P. Cotter, P.~F. Griffin, E.~A. Hinds, C.~N.
  Ironside, P.~See, A.~G. Sinclair, E.~Riis, and A.~S. Arnold, \enquote{{A
  surface-patterned chip as a strong source of ultracold atoms for quantum
  technologies},} {\protect\JournalTitle{Nature Nanotechnology}} \textbf{8},
  321 (2013).

\bibitem{Cotter2016}
J.~P. Cotter, J.~P. McGilligan, P.~F. Griffin, I.~M. Rabey, K.~Docherty,
  E.~Riis, A.~S. Arnold, and E.~A. Hinds, \enquote{{Design and fabrication of
  diffractive atom chips for laser cooling and trapping},}
  {\protect\JournalTitle{Applied Physics B}} \textbf{122}, 172 (2016).

\bibitem{McGehee2021}
W.~R. McGehee, W.~Zhu, D.~S. Barker, D.~Westly, A.~Yulaev, N.~Klimov,
  A.~Agrawal, S.~Eckel, V.~Aksyuk, and J.~J. McClelland,
  \enquote{{Magneto-optical trapping using planar optics},}
  {\protect\JournalTitle{New Journal of Physics}} \textbf{23}, 013021 (2021).

\bibitem{McGilligan2020}
J.~P. McGilligan, K.~R. Moore, A.~Dellis, G.~D. Martinez, E.~{De Clercq}, P.~F.
  Griffin, A.~S. Arnold, E.~Riis, R.~Boudot, and J.~Kitching, \enquote{{Laser
  cooling in a chip-scale platform},} {\protect\JournalTitle{Applied Physics
  Letters}} \textbf{117}, 054001 (2020).

\bibitem{Burrow2021}
O.~S. Burrow, P.~F. Osborn, E.~Boughton, F.~Mirando, D.~P. Burt, A.~S. Arnold,
  P.~F. Griffin, and E.~Riis, \enquote{{A centilitre-scale vacuum chamber for
  compact ultracold quantum technologies},} ArXiv:2101.07851 (2021).

\bibitem{Raab1987}
E.~L. Raab, M.~Prentiss, A.~Cable, S.~Chu, and D.~E. Pritchard,
  \enquote{{Trapping of Neutral Sodium Atoms with Radiation Pressure},}
  {\protect\JournalTitle{Physical Review Letters}} \textbf{59}, 2631 (1987).

\bibitem{McGilligan2015}
J.~P. McGilligan, P.~F. Griffin, E.~Riis, and A.~S. Arnold,
  \enquote{{Phase-space properties of magneto-optical traps utilising
  micro-fabricated gratings},} {\protect\JournalTitle{Opt. Express}}
  \textbf{23}, 8948 (2015).

\bibitem{Imhof2017}
E.~Imhof, B.~K. Stuhl, B.~Kasch, B.~Kroese, S.~E. Olson, and M.~B. Squires,
  \enquote{{Two-dimensional grating magneto-optical trap},}
  {\protect\JournalTitle{Physical Review A}} \textbf{96}, 033636 (2017).

\bibitem{Ungar1989}
P.~J. Ungar, D.~S. Weiss, E.~Riis, and S.~Chu, \enquote{{Optical molasses and
  multilevel atoms: theory},} {\protect\JournalTitle{Journal of the Optical
  Society of America B}} \textbf{6}, 2058 (1989).

\bibitem{Lett1989}
P.~D. Lett, W.~D. Phillips, S.~L. Rolston, C.~E. Tanner, R.~N. Watts, and C.~I.
  Westbrook, \enquote{{Optical molasses},} {\protect\JournalTitle{Journal of
  the Optical Society of America B}} \textbf{6}, 2084 (1989).

\bibitem{Petsas1994}
K.~I. Petsas, A.~B. Coates, and G.~Grynberg, \enquote{{Crystallography of
  optical lattices},} {\protect\JournalTitle{Physical Review A}} \textbf{50},
  5173 (1994).

\bibitem{Lee2013a}
J.~Lee, J.~A. Grover, L.~A. Orozco, and S.~L. Rolston, \enquote{{Sub-Doppler
  cooling of neutral atoms in a grating magneto-optical trap},}
  {\protect\JournalTitle{J. Opt. Soc. Am. B}} \textbf{30}, 2869 (2013).

\bibitem{Boiron1995}
D.~Boiron, C.~Trich{\'{e}}, D.~R. Meacher, P.~Verkerk, and G.~Grynberg,
  \enquote{{Three-dimensional cooling of cesium atoms in four-beam gray optical
  molasses},} {\protect\JournalTitle{Physical Review A}} \textbf{52}, R3425
  (1995).

\bibitem{Hauth2013}
M.~Hauth, C.~Freier, V.~Schkolnik, A.~Senger, M.~Schmidt, and A.~Peters,
  \enquote{{First gravity measurements using the mobile atom interferometer
  GAIN},} {\protect\JournalTitle{Applied Physics B}} \textbf{113}, 49 (2013).

\bibitem{Wu2017}
X.~Wu, F.~Zi, J.~Dudley, R.~J. Bilotta, P.~Canoza, and H.~M{\"{u}}ller,
  \enquote{{Multiaxis atom interferometry with a single diode laser and a
  pyramidal magneto-optical trap},} {\protect\JournalTitle{Optica}} \textbf{4},
  1545 (2017).

\bibitem{Brown2019}
M.~O. Brown, T.~Thiele, C.~Kiehl, T.~W. Hsu, and C.~A. Regal,
  \enquote{{Gray-Molasses Optical-Tweezer Loading: Controlling Collisions for
  Scaling Atom-Array Assembly},} {\protect\JournalTitle{Physical Review X}}
  \textbf{9}, 011057 (2019).

\bibitem{Sitaram2020}
A.~Sitaram, P.~K. Elgee, G.~K. Campbell, N.~N. Klimov, S.~Eckel, and D.~S.
  Barker, \enquote{{Confinement of an alkaline-earth element in a grating
  magneto-optical trap},} {\protect\JournalTitle{Review of Scientific
  Instruments}} \textbf{91}, 103202 (2020).

\bibitem{Valentin1992}
C.~Valentin, M.~C. Gagn{\'{e}}, J.~Yu, and P.~Pillet, \enquote{{One-dimension
  sub-doppler molasses in the presence of static magnetic field},}
  {\protect\JournalTitle{EPL (Europhysics Letters)}} \textbf{17}, 133 (1992).

\bibitem{Grynberg1994}
G.~Grynberg and J.-Y. Courtois, \enquote{{Proposal for a magneto-optical
  lattice for trapping atoms in nearly-dark states},}
  {\protect\JournalTitle{EPL (Europhysics Letters)}} \textbf{27}, 41 (1994).

\bibitem{Shahriar1993}
M.~S. Shahriar, P.~R. Hemmer, M.~G. Prentiss, P.~Marte, J.~Mervis, D.~P. Katz,
  N.~P. Bigelow, and T.~Cai, \enquote{{Continuous polarization-gradient
  precooling-assisted velocity-selective coherent population trapping},}
  {\protect\JournalTitle{Physical Review A}} \textbf{48}, R4035 (1993).

\bibitem{Marte1994}
P.~Marte, R.~Dum, R.~Ta{\"{i}}eb, P.~Zoller, M.~S. Shahriar, and M.~Prentiss,
  \enquote{{Polarization-gradient-assisted subrecoil cooling: Quantum
  calculations in one dimension},} {\protect\JournalTitle{Physical Review A}}
  \textbf{49}, 4826 (1994).

\bibitem{Weidemuller1994}
M.~Weidem{\"{u}}ller, T.~Esslinger, M.~A. Ol'shanii, A.~Hemmerich, and T.~W.
  H{\"{a}}nsch, \enquote{{A Novel Scheme for Efficient Cooling below the Photon
  Recoil Limit},} {\protect\JournalTitle{Europhysics Letters}} \textbf{27}, 109
  (1994).

\bibitem{Aspect1988}
A.~Aspect, E.~Arimondo, R.~Kaiser, N.~Vansteenkiste, and C.~Cohen-Tannoudji,
  \enquote{{Laser Cooling below the One-Photon Recoil Energy by
  Velocity-Selective Coherent Population Trapping},}
  {\protect\JournalTitle{Physical Review Letters}} \textbf{61}, 826 (1988).

\bibitem{Fernandes2012}
D.~R. Fernandes, F.~Sievers, N.~Kretzschmar, S.~Wu, C.~Salomon, and F.~Chevy,
  \enquote{{Sub-Doppler laser cooling of fermionic 40 K atoms in
  three-dimensional gray optical molasses},} {\protect\JournalTitle{EPL
  (Europhysics Letters)}} \textbf{100}, 63001 (2012).

\bibitem{Grier2013}
A.~T. Grier, I.~Ferrier-Barbut, B.~S. Rem, M.~Delehaye, L.~Khaykovich,
  F.~Chevy, and C.~Salomon, \enquote{{$\Lambda$-enhanced sub-Doppler cooling of
  lithium atoms in $D_1$ gray molasses},} {\protect\JournalTitle{Physical
  Review A}} \textbf{87}, 063411 (2013).

\bibitem{Truppe2017}
S.~Truppe, H.~J. Williams, M.~Hambach, L.~Caldwell, N.~J. Fitch, E.~A. Hinds,
  B.~E. Sauer, and M.~R. Tarbutt, \enquote{{Molecules cooled below the Doppler
  limit},} {\protect\JournalTitle{Nature Physics}} \textbf{13}, 1173 (2017).

\bibitem{Anderegg2018}
L.~Anderegg, B.~L. Augenbraun, Y.~Bao, S.~Burchesky, L.~Cheuk, W.~Ketterle, and
  J.~M. Doyle, \enquote{{Laser cooling of optically trapped molecules},}
  {\protect\JournalTitle{Nature Physics}} \textbf{14}, 890 (2018).

\bibitem{McCarron2018}
D.~J. McCarron, M.~H. Steinecker, Y.~Zhu, and D.~Demille, \enquote{{Magnetic
  Trapping of an Ultracold Gas of Polar Molecules},}
  {\protect\JournalTitle{Physical Review Letters}} \textbf{121}, 013202 (2018).

\bibitem{Rosi2018}
S.~Rosi, A.~Burchianti, S.~Conclave, D.~S. Naik, G.~Roati, C.~Fort, and
  F.~Minardi, \enquote{{$\Lambda$-enhanced grey molasses on the $D_2$
  transition of Rubidium-87 atoms},} {\protect\JournalTitle{Scientific
  Reports}} \textbf{8}, 1301 (2018).

\bibitem{Ornelas-Huerta2020}
D.~P. Ornelas-Huerta, A.~N. Craddock, E.~A. Goldschmidt, A.~J. Hachtel,
  Y.~Wang, P.~Bienias, A.~V. Gorshkov, S.~L. Rolston, and J.~V. Porto,
  \enquote{{On-demand indistinguishable single photons from an efficient and
  pure source based on a Rydberg ensemble},} {\protect\JournalTitle{Optica}}
  \textbf{7}, 813 (2020).

\bibitem{Grunzweig2010}
T.~Gr{\"{u}}nzweig, A.~Hilliard, M.~McGovern, and M.~F. Andersen,
  \enquote{{Near-deterministic preparation of a single atom in an optical
  microtrap},} {\protect\JournalTitle{Nature Physics}} \textbf{6}, 951 (2010).

\bibitem{Starkey2013}
P.~T. Starkey, C.~J. Billington, S.~P. Johnstone, M.~Jasperse, K.~Helmerson,
  L.~D. Turner, and R.~P. Anderson, \enquote{{A scripted control system for
  autonomous hardware-timed experiments},} {\protect\JournalTitle{Review of
  Scientific Instruments}} \textbf{84}, 085111 (2013).

\bibitem{Norrgard2018}
E.~B. Norrgard, D.~S. Barker, J.~A. Fedchak, N.~Klimov, J.~Scherschligt, and
  S.~P. Eckel, \enquote{{Note: A 3D-printed alkali metal dispenser},}
  {\protect\JournalTitle{Review of Scientific Instruments}} \textbf{89}, 056101
  (2018).

\bibitem{Fedchak2018}
J.~A. Fedchak, J.~Scherschligt, D.~Barker, S.~Eckel, A.~P. Farrell, and
  M.~Sefa, \enquote{{Vacuum furnace for degassing stainless-steel vacuum
  components},} {\protect\JournalTitle{Journal of Vacuum Science and Technology
  A}} \textbf{36}, 023201 (2018).

\bibitem{Barker2019a}
D.~S. Barker, N.~C. Pisenti, A.~Restelli, J.~Scherschligt, J.~A. Fedchak, G.~K.
  Campbell, and S.~Eckel, \enquote{{A flexible, open-source radio-frequency
  driver for acousto-optic and electro-optic devices},} arXiv:1908.02156
  (2019).

\bibitem{Phillips1982}
W.~D. Phillips and H.~Metcalf, \enquote{{Laser deceleration of an atomic
  beam},} {\protect\JournalTitle{Physical Review Letters}} \textbf{48}, 596
  (1982).

\bibitem{gratingnote}
We use the same grating chip as Ref.~\cite{Barker2019}, but we have corrected
  the normalization of the Stokes parameters and the sign of \(V\).

\bibitem{Shimizu1991}
F.~Shimizu, K.~Shimizu, and H.~Takuma, \enquote{{Four-beam laser trap of
  neutral atoms},} {\protect\JournalTitle{Optics Letters}} \textbf{16}, 339
  (1991).

\bibitem{Lin1991}
Z.~Lin, K.~Shimizu, M.~Zhan, F.~Shimizu, and H.~Takuma, \enquote{{Laser Cooling
  and Trapping of Li},} {\protect\JournalTitle{Japanese Journal of Applied
  Physics}} \textbf{30}, L 1324 (1991).

\bibitem{Burchianti2014}
A.~Burchianti, G.~Valtolina, J.~A. Seman, E.~Pace, M.~{De Pas}, M.~Inguscio,
  M.~Zaccanti, and G.~Roati, \enquote{{Efficient all-optical production of
  large $^6$Li quantum gases using $D_1$ gray-molasses cooling},}
  {\protect\JournalTitle{Physical Review A}} \textbf{90}, 043408 (2014).

\bibitem{Sievers2015}
F.~Sievers, N.~Kretzschmar, D.~R. Fernandes, D.~Suchet, M.~Rabinovic, S.~Wu,
  C.~V. Parker, L.~Khaykovich, C.~Salomon, and F.~Chevy, \enquote{{Simultaneous
  sub-Doppler laser cooling of fermionic $^6$Li and $^{40}$K on the $D_1$ line:
  Theory and experiment},} {\protect\JournalTitle{Physical Review A}}
  \textbf{91}, 023426 (2015).

\bibitem{Satter2018}
C.~L. Satter, S.~Tan, and K.~Dieckmann, \enquote{{Comparison of an efficient
  implementation of gray molasses to narrow-line cooling for the all-optical
  production of a lithium quantum gas},} {\protect\JournalTitle{Physical Review
  A}} \textbf{98}, 023422 (2018).

\bibitem{Kim2019}
K.~Kim, S.~Huh, K.~Kwon, and J.~Choi, \enquote{{Rapid production of large
  $^7$Li Bose-Einstein condensates using $D_1$ gray molasses},}
  {\protect\JournalTitle{Physical Review A}} \textbf{99}, 053604 (2019).

\bibitem{Salomon2013}
G.~Salomon, L.~Fouch{\'{e}}, P.~Wang, A.~Aspect, P.~Bouyer, and T.~Bourdel,
  \enquote{{Gray-molasses cooling of 39K to a high phase-space density},}
  {\protect\JournalTitle{EPL (Europhysics Letters)}} \textbf{104}, 63002
  (2013).

\bibitem{Nath2013}
D.~Nath, R.~K. Easwaran, G.~Rajalakshmi, and C.~S. Unnikrishnan,
  \enquote{{Quantum-interference-enhanced deep sub-Doppler cooling of $^{39}$K
  atoms in gray molasses},} {\protect\JournalTitle{Physical Review A}}
  \textbf{88}, 053407 (2013).

\bibitem{Bouton2015}
Q.~Bouton, R.~Chang, A.~L. Hoendervanger, F.~Nogrette, A.~Aspect, C.~I.
  Westbrook, and D.~Cl{\'{e}}ment, \enquote{{Fast production of Bose-Einstein
  condensates of metastable helium},} {\protect\JournalTitle{Physical Review
  A}} \textbf{91}, 061402(R) (2015).

\bibitem{Chen2016}
H.-Z. Chen, X.-C. Yao, Y.-P. Wu, X.-P. Liu, X.-Q. Wang, Y.~X. Wang, Y.-A. Chen,
  and J.-W. Pan, \enquote{{Production of large $^{41}$K Bose-Einstein
  condensates using $D_1$ gray molasses},} {\protect\JournalTitle{Physical
  Review A}} \textbf{94}, 033408 (2016).

\bibitem{Colzi2016}
G.~Colzi, G.~Durastante, E.~Fava, S.~Serafini, G.~Lamporesi, and G.~Ferrari,
  \enquote{{Sub-Doppler cooling of sodium atoms in gray molasses},}
  {\protect\JournalTitle{Physical Review A}} \textbf{93}, 023421 (2016).

\bibitem{Eckel2020}
S.~Eckel, D.~S. Barker, E.~B. Norrgard, and J.~Scherschligt, \enquote{{PyLCP: A
  python package for computing laser cooling physics},} ArXiv:2011.07979
  (2020).

\bibitem{polnote}
We expect that the effect of polarization impurity due to the grating Stokes
  parameters is small compared to the effect of projecting the polarization on
  to the quantization axis \(\hat{z}\).

\bibitem{Ashkin1980}
A.~Ashkin and J.~P. Gordon, \enquote{{Motion of atoms in a radiation trap},}
  {\protect\JournalTitle{Physical Review A}} \textbf{21}, 1606 (1980).

\bibitem{Bruce2017}
G.~D. Bruce, E.~Haller, B.~Peaudecerf, D.~A. Cotta, M.~Andia, S.~Wu, M.~Y.~H.
  Johnson, B.~W. Lovett, and S.~Kuhr, \enquote{{Sub-Doppler laser cooling of
  $^{40}$K with Raman gray molasses on the $D_2$ line},}
  {\protect\JournalTitle{Journal of Physics B: Atomic, Molecular and Optical
  Physics}} \textbf{50}, 095002 (2017).

\bibitem{techsimnote}
We have also checked that applying the magnetic field along a radial axis or
  modelling the Raman detuning jitter as a Gaussian-distributed random process
  produces similar simulation results.

\bibitem{Knuth2019}
K.~H. Knuth, \enquote{{Optimal data-based binning for histograms and
  histogram-based probability density models},} {\protect\JournalTitle{Digital
  Signal Processing}} \textbf{95}, 102581 (2019).

\bibitem{Kass1995}
K.~E. Kass and A.~E. Raftery, \enquote{{Bayes Factors},}
  {\protect\JournalTitle{Journal of the American Statistical Association}}
  \textbf{90}, 773 (1995).

\bibitem{Burnham2004}
K.~P. Burnham and D.~R. Anderson, \enquote{{Multimodel inference: Understanding
  AIC and BIC in model selection},} {\protect\JournalTitle{Sociological Methods
  and Research}} \textbf{33}, 261 (2004).

\bibitem{bicnote}
We use the Bayesian information criterion, rather than alternative model
  selectors like \(\chi^{2}\) or the Akaike information criterion, because it
  imposes the largest penalty on models with more free parameters.

\end{thebibliography}

\end{document}